\documentclass{sobolev} % Please, don't modify!

\setpg{1} % Please, don't modify!

\titl[A viscous-convective instability in laminar Keplerian thin discs]{A viscous-convective instability in laminar Keplerian thin discs} 

\authors{K.L.\,Malanchev et al.}{K.L.\,Malanchev\aff{1,2}, K.A.\,Postnov\aff{1,2}, N.I.\,Shakura\aff{1}}

\affiliat{
 \item M.V. Lomonosov Moscow State University, Sternberg Astronomical Institute, Universitetsky pr., 13, Moscow, 119234, Russia   
 \item M.V. Lomonosov Moscow State University, Faculty of Physics, Leninskie Gory, GSP-1, Moscow, 119991, Russia
}

\email{malanchev@physics.msu.ru} % Contact e-mail

\usepackage{bm}

\begin{document} 

\begin{abstract}
Using the anelastic approximation of linearized hydrodynamic equations, we investigate the development of axially symmetric small perturbations in thin Keplerian discs.
Dispersion relation is found as a solution of general Sturm--Liouville eigenvalue problem for different values of relevant physical parameters (viscosity, heat conductivity, disc semithickness).
The analysis reveals the appearance of overstable mode for Prandtl parameter higher than some critical value.
These modes have a viscous-convective nature and can serve as a seed for turbulence in astrophysical discs even in the absence of magnetic fields.
\end{abstract}  

\section{Introduction}
The problem of linear stability of sheared astrophysical flows has been actively studied. The recent papers~\cite{shakura_postnov2015a},~\cite{shakura_postnov2015b} 
used the Boussinesq and anelastic approximations, respectively, with taking into account microscopic viscosity and thermal conductivity of the gas.
These analyses have revealed the presence of overstable viscous modes whose physical origin is likely to be connected to development of convective motions in vertically stratified accretion flows.
However, in those papers averaging over vertical disc structure was performed, which restricted applications of the obtained results. In the present paper, we take into account more realistic polytropic structure of a Keplerian accretion disc and solve linearized general Sturm--Liuville eigenvalue problem.
Our analysis confirms the appearance of the overstable modes in the wide range of microscopic parameters of the gas described by the Prandtl number.

\section{Basic equations}
The system of hydrodynamic equations for axially symmetric accretion flow can be written as follows

\begin{enumerate}
 \item Continuity equation:
  \begin{equation}
   \frac{\partial \rho}{\partial t}+\frac{1}{r}\frac{\partial (\rho r u_r) }{\partial r}
+\frac{\partial (\rho u_z) }{\partial z}=0\,.
  \label{eq.C}
  \end{equation}
  
  The anelastic approximation for gas velocity is $\bm{u}$ is $\nabla \cdot \rho_0\bm{u}$ = 0.
 
 \item The radial, azimuthal and vertical components of the Navier--Stokes momentum equation are, respectively:
  \begin{equation}
   \frac{\partial u_r}{\partial t}+ u_r \frac{\partial u_r}{\partial r}+
u_z\frac{\partial u_r}{\partial z}-\frac{u_\phi^2}{r}=
-\frac{\partial \phi_g}{\partial r}-\frac{1}{\rho}\frac{\partial p}{\partial r}+{\cal N}_r\,,
  \label{eq.R}
  \end{equation}
  
  \begin{equation}
   \frac{\partial u_\phi}{\partial t}+ u_r \frac{\partial u_\phi}{\partial r}+
u_z\frac{\partial u_\phi}{\partial z}+\frac{u_r u_\phi}{r}={\cal N}_\phi\,,
  \label{eq.PHI}
  \end{equation}
  
  \begin{equation}
   \frac{\partial u_z}{\partial t}+ u_r \frac{\partial u_z}{\partial r}+
u_z\frac{\partial u_z}{\partial z}=
-\frac{\partial \phi_g}{\partial z}-\frac{1}{\rho}\frac{\partial p}{\partial z}+{\cal N}_z\,,
  \label{eq.Z}
  \end{equation}
where ${\cal N}_r$, ${\cal N}_\phi$ and ${\cal N}_z$ are viscous forces. For their specific form see for instance~\cite{kato}.

  In this work we will drop the second derivatives of velocities $u$ with respect to the vertical coordinate $z$ in the Navier--Stokes equations following~\cite{shakura_postnov2015b}. This assumption makes the problem simpler and the more general problem will be solved in~\cite{malanchev_etal2016}.

 \item Energy equation:
  \begin{equation}
   \frac{\rho {\cal R} T}{\mu}\left[\frac{\partial s}{\partial t}+(\bm{u}\nabla)\cdot s\right] = Q_\mathrm{visc}-\nabla\cdot\bm{F}\,,
  \label{eq.E}
  \end{equation}
  where $s$ is specific entropy per particle, $Q_\mathrm{visc}$ is the viscous dissipation rate per unit volume, R is the universal gas constant, $\mu$ is the molecular weight, $T$ is the temperature and terms on the right stand for the viscous energy production and the heat conductivity energy flux $F$, respectively. The energy flux due to the heat conductivity is
  \begin{equation}
   \nabla\cdot\bm{F}=\nabla(-\kappa\nabla T)=-\kappa\Delta T-\nabla\kappa\cdot\nabla T\,.
  \label{eq.energy_vlux}
  \end{equation}

  In the Boussinesq approximation, in the energy equation the Eulerian perturbations should be zero: $p_1 = 0$. Following \cite{shakura_postnov2015a}, we will also drop the term $\nabla\kappa\cdot\nabla T$ but keep $\kappa\Delta T$.
\end{enumerate}

\section{Linearized equations in the anelastic approximation}
Perturbed hydrodynamic variables can be written in the form $x = x_0 + x_1$, where $x_0$ stands for the unperturbed background quantities and $x_1 = (\rho_1, p_1, u_{r\,1}, u_{z\,1}, u_{\phi\, 1})$ are small perturbations. We take all these small perturbations in the form $x_1 = f(z) \cdot \exp{(\mathrm{i} \omega t - \mathrm{i} k_r r)}$. We will consider thin discs with semithickness $z_0 / r \ll 1$ and relatively large radial wavenumbers of perturbations $k_r r \gg 1$. Small thickness of the disc and large wavenumbers allow us to set the radial derivatives to zero, $\partial x_0 / \partial r = 0$. Under these assumptions, linearizing of the system of equations~(\ref{eq.C}---\ref{eq.E}) yields the following system of equations\cite{shakura_postnov2015a, shakura_postnov2015b}:
\begin{enumerate}
 \item Continuity equation
  \begin{equation}
   \frac{\partial u_z}{\partial z}-\mathrm{i}k_ru_r+\frac1{\rho_0}\frac{\partial \rho_0}{\partial z}u_z=0.
  \end{equation}
 
 \item Momentum equations
  \begin{equation}
   (\mathrm{i}\omega +\nu k_r^2)u_r-2\Omega u_\phi=\mathrm{i}k_r\frac{p_1}{\rho_0}\,,
  \end{equation}
  \begin{equation}
   (\mathrm{i}\omega+\nu k_r^2)u_\phi+\frac{\varkappa^2}{2\Omega}u_r=0\,,
  \end{equation}
  \begin{equation}
   (\mathrm{i}\omega +\nu k_r^2)u_z=-\frac1{\rho_0}\frac{\partial p_1}{\partial z}+\frac{\rho_1}{\rho_0} \frac1{\rho_0}\frac{\partial p_0}{\partial z}.
  \end{equation}

 \item Energy equation
  \begin{equation}
   \begin{split}
    \frac{\rho_1}{\rho_0} \left[ \mathrm{i}\omega  + \frac{\nu k_r^2}{\mathrm{Pr}} - \alpha_\mathrm{visc}\frac{\nu}{\mathrm{Pr}}\frac1{T_0}\frac{\partial^2 T_0}{\partial z^2} - \alpha_\mathrm{visc}\nu \left(r\frac{d\Omega}{dr}\right)^2 \frac{\mu}{{\cal R}T_0} \right]&\\
    = \frac{2\mathrm{i}k_r \nu r (d \Omega/d r)}{c_p{\cal R}T_0/\mu}u_{\phi}+\frac{1}{c_p}\frac{\partial s_0}{\partial z}u_z&,
   \end{split}
  \label{eq.E1}
  \end{equation}
  where $\mathrm{Pr}$ is the Prandtl number, ${\cal R}$ is the universal gas constant, $\varkappa$ is the epicyclic frequency, $\nu \sim T_0^{\alpha_\mathrm{visc}} / \rho_0$ is the kinematic viscosity coefficient. The kinematic viscosity in the disc equatorial plane is $\nu|_{z=0} = (v_s/v_\phi)\,(l/r) \Omega r^2$, where $v_s$ is the sound velocity and $l$ is the mean free path of particles. We assume the gas to be fully ionized so that $\alpha_\mathrm{visc} = 5/2$.
\end{enumerate}

It is necessary to set the background solution of hydrodynamic equations to find solution for perturbations.
As the background state, we will use adiabatic polytropic discs~\cite{ketsaris_shakura1998}:
\begin{eqnarray*}
 &T_0(z) = T_\mathrm{c} \left( 1 - \left(\frac{z}{z_0}\right)^2 \right)\,,
 &\rho_0(z) = \rho_\mathrm{c} \left( 1 - \left(\frac{z}{z_0}\right)^2 \right)^{3/2}\,,\\
 &p_0(z) = p_\mathrm{c} \left( 1 - \left(\frac{z}{z_0}\right)^2 \right)^{5/2}\,,
 &s_0(z) = c_p \log{\left( \frac{p_0^{3/5}}{\rho_0} \right)} = const.
\end{eqnarray*}

This system of algebraic and differential equations can be transformed to one second-order differential equation for pressure perturbations $p_1$:
%These system of equations provides a way to find dispersion relation.
%For every value of wavenumber $k_r$ one should solve general Sturm–Liouville problem for eigenvalue $\omega$.

\begin{equation}
 \frac{\partial^2 p_1}{\partial z} - \alpha(\omega) \frac{z}{z_0^2} \frac{\partial p_1}{\partial z} - (\alpha(\omega)+\beta(\omega)) \frac1{z_0^2} p_1 = 0\,,
\label{eq.p1z}
\end{equation}
where $z_0 = \sqrt{3}\, (v_s/v_\phi)\, r$ is the disc semithickness~\cite{ketsaris_shakura1998}, and the dimensionless coefficients $\alpha$ and $\beta$ reads:

\begin{equation}
 \begin{split}
  \alpha(\omega) &= \left(\frac{v_s}{v_\phi}\right) \left(\frac{l}{r}\right) \dfrac{\varkappa^2}{(\mathrm{i}\omega+\nu k_r^2)^2} \,\times\\
  &\times\, \dfrac{(- d\ln\Omega/d\ln r)}{c_p\left[ \mathrm{i}\omega + \dfrac{\nu k_r^2}{\mathrm{Pr}} + 2\dfrac{\alpha_\mathrm{visc}}{z_0^2}\dfrac{\nu}{T_0} - \alpha_\mathrm{visc}\nu\left(r\dfrac{d \Omega}{d r}\right)^2\dfrac{\mu}{{\cal R}T_0} \right]} \dfrac{(k_r r)^2}{1+\dfrac{\varkappa^2}{(\mathrm{i}\omega+\nu k_r^2)^2}}\,,
 \end{split}
\end{equation}

\begin{equation}
 \beta(\omega) = \dfrac{(k_r r)^2}{1+\dfrac{\varkappa^2}{(\mathrm{i}\omega+\nu k_r^2)^2}}\,,
\end{equation}

The pressure perturbation $p_1$ must vanish at the disc boundary  ($z = z_0$), and the function $p_1(z)$ should be even or odd because of the plane symmetry of the problem.
Bellow we will consider the case of even $p_1(z)$ with the boundary conditions:
\begin{equation}
 \left. \frac{\partial p_1}{\partial z} \right|_{z = 0} = 0\,,
\label{eq.p1z_0}
\end{equation}
\begin{equation}
 \left. p_1 \right|_{z = z_0} = 0\,.
\label{eq.p1z_z0}
\end{equation}
We are searching for the least oscillating solutions, which means that $p_1(z)$ should not have zeros between $z = 0$ and $z = z_0$.
This condition comes from our previous assumption about the smallness of the secondary derivatives of velocities in the Navier--Stokes equations~(\ref{eq.R}--\ref{eq.Z}).

Using a new variable $x \equiv z/z_0$, equation (\ref{eq.p1z}) transforms to:
\begin{equation}
	\frac{\partial^2 p_1}{\partial x^2} - \alpha(\omega)\, \frac{\partial p_1}{\partial x} - (\alpha(\omega)+\beta(\omega)) p_1 = 0.
\label{eq.p1x}
\end{equation}

Let us introduce a new function $Y(x)$:
\begin{equation}
	Y \equiv p_1 \times \exp\left( \frac12 \int_0^x{ -\alpha(\omega) x' d x' } \right) = p_1 \times \exp \left( - \frac{\alpha(\omega)\,x^2}{4} \right).
\end{equation}

Then \eqref{eq.p1x} transforms to:
\begin{eqnarray}
	\frac{\partial^2 Y}{\partial x^2} + \left( -\beta(\omega) - \frac{\alpha(\omega)}{2} - \frac{\alpha^2(\omega)\,x^2}{4} \right) Y = 0.
\end{eqnarray}

After introducing the new variable $\zeta \equiv \sqrt{\alpha(\omega)}\,x$, the eigenvalue problem~(\ref{eq.p1z},~\ref{eq.p1z_0},~\ref{eq.p1z_z0}) takes the form:
\begin{eqnarray}
	&\displaystyle \frac{\partial^2 Y}{\partial \zeta^2} - \left( \frac{\zeta^2}{4} + \left[ \frac{\beta(\omega)}{\alpha(\omega)} + \frac12 \right] \right) Y = 0,\label{eq.canonic}\\
	&\displaystyle \left. \frac{\partial Y}{\partial \zeta} \right|_{\zeta = 0} = 0,\label{eq.bound_zeta_0}\\
	&\displaystyle Y|_{\zeta = \sqrt{\alpha(\omega)}} = 0.\label{eq.bound_zeta_z0}
\end{eqnarray}

Our boundary condition \eqref{eq.bound_zeta_0} and the plane symmetry of the problem enable  us to use an even solution of equation \eqref{eq.canonic}~\cite{abramowitz_stegun}:
\begin{equation}
 \begin{split}
	Y_\eta(\zeta) &= e^{-\zeta^2/4} \times M( -\frac{\eta}{2}, \frac12, \frac{\zeta^2}{2} ) =\\
		&= e^{-\zeta^2/4} \left\{ 1 + (-\eta)\frac{\zeta^2}{2!} + (-\eta)(-\eta+2)\frac{\zeta^4}{4!} + ... \right\},
 \end{split}
\label{eq.even_eigen}
\end{equation}
where $M$ is the confluent hypergeometric function, $\eta(\omega) \equiv -1 - \beta(\omega)/\alpha(\omega)$.

An eigenfunction of the problem (with the boundary condition \eqref{eq.bound_zeta_z0}) must satisfy the following relation:
\begin{equation}
	Y_{\eta(\omega)}(\sqrt{\alpha(\omega)}) = 0.
\end{equation}
The last relation can be regarded as an equation for the unknown variable $\omega$, 
then the solutions to this equations are eigenvalues of our problem, and the corresponding $Y_\eta(\zeta)$ will be its eigenfunctions.

\section{Dispersion relation}
\begin{figure}[h]
  \centering
  \includegraphics[width=0.7\paperwidth]{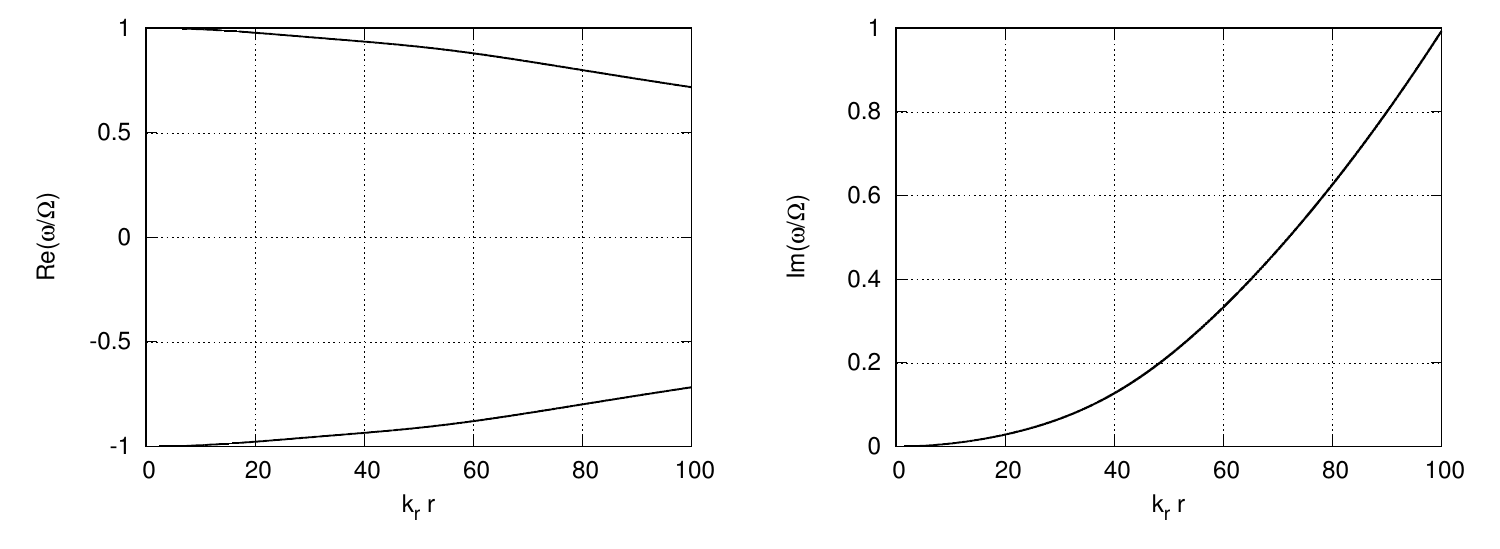}
  \caption{The dispersion equation for the critical Prandtl number $\mathrm{Pr} = 8/45$, the mean-free path length of particles $l/r = 0.01$ and disc semi-thickness parameter $v_s/v_\phi = 0.01$. The left panel shows the real part of two decaying modes in terms of dimensionless frequency $\omega/\Omega$ and the dimensionless wavenumber $k_r r$. The right panel shows the imaginary part of the dispersion relation which is the same for both decaying modes. }
  \label{fig.decaying}
\end{figure}

\begin{figure}[h]
  \centering
  \includegraphics[width=0.7\paperwidth]{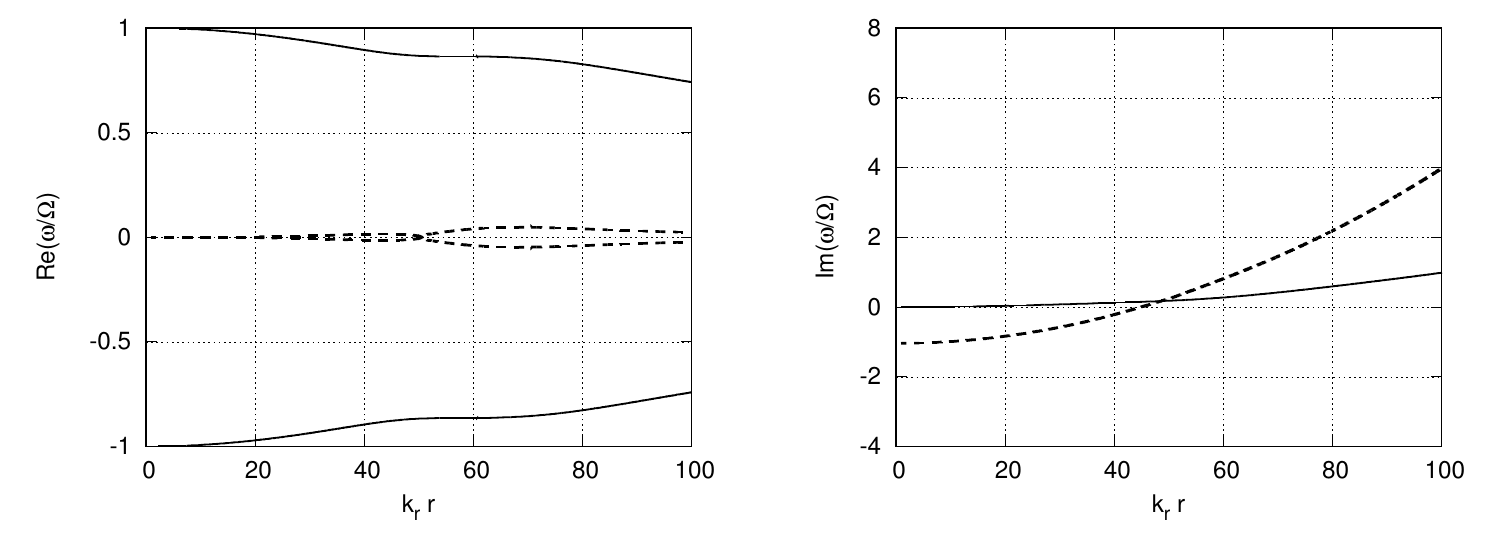}
  \caption{Dispersion relation for $\mathrm{Pr} = 0.2$, $l/r = 0.01$ and $v_s/v_\phi = 0.01$. Left panel shows the real part of two decaying (the solid line) and to the overstable modes (the dashed line) in terms of the dimensionless frequency $\omega/\Omega$ and the dimensionless wavenumber $k_r r$. Right panels shows the imaginary part of these modes. }
  \label{fig.overstable}
\end{figure}

Solution of the eigenvalue problem depends of the sign of the linearized term $\kappa \Delta T$, which appears in the energy equation~\eqref{eq.E1} in the form:
\begin{equation}
\alpha_\mathrm{visc}\frac{\nu}{\mathrm{Pr}}\frac1{T_0}\frac{\partial^2 T_0}{\partial z^2} + \alpha_\mathrm{visc}\nu \left(r\frac{d\Omega}{dr}\right)^2 \frac{\mu}{{\cal R}T_0}\,.
\label{eq.termPr}
\end{equation}
If the Prandtl number $\mathrm{Pr} \le 8/45$ and this relation is negative, there is only one mode of the dispersion equation which corresponds to a decaying mode (see Fig.~\ref{fig.decaying}).
Otherwise, if $\mathrm{Pr} > 8/45$, an additional overstable mode appears (see Fig.~\ref{fig.overstable}).

\begin{figure}[h]
  \centering
  \includegraphics[width=0.7\paperwidth]{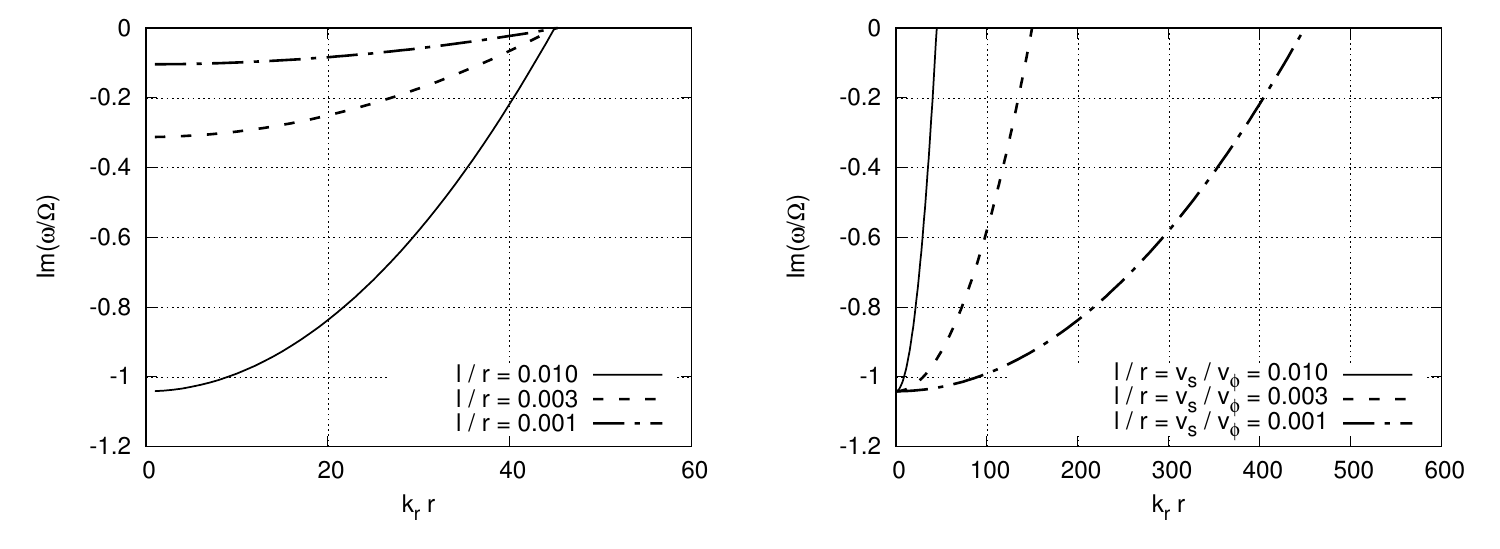}
  \caption{The imaginary part of the dispersion relations for $\mathrm{Pr} = 0.2$ and different values of $l/r$ and $v_s/v_\phi$. In the left panel, the value of $v_s/v_\phi = 0.01$ is constant for all curves. The viscosity changes in proportion to $l/r$ and $\mathrm{Im}(\omega)$ changes in the same way. On the right panel both $v_s/v_\phi$ and $l/r$ changes in the same way so that the term~\eqref{eq.termPr} keeps constant. Here the range of wavenumbers $k_r r$ of the overstable mode decreases inversely proportional to the disc thickness. }
  \label{fig.varl_varv}
\end{figure}

\begin{figure}[h]
  \centering
  \includegraphics[width=0.45\paperwidth]{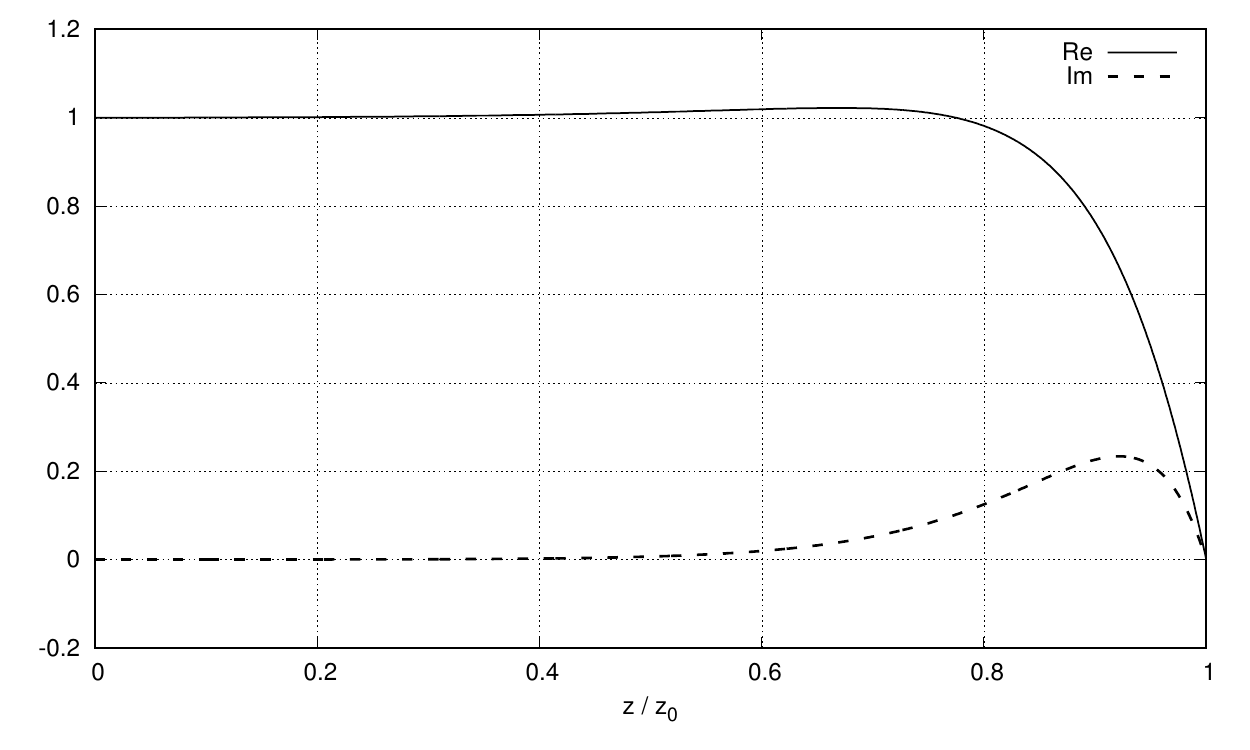}
  \caption{The overstable solution of the problem~(\ref{eq.p1z},~\ref{eq.p1z_0},~\ref{eq.p1z_z0}) for variable $x = z/z_0$ with the following parameters: $k_r r = 40$, $\mathrm{Pr} = 0.2$, $l/r = 0.01$ and $v_s/v_\phi = 0.01$. Figure shows the normalized eigenfunction $p_1(x)$. The solid line shows the real part of $p_1(x)$, the dashed line shows the imaginary part of the pressure perturbation $p_1(x)$.}
  \label{fig.eigenfunction}
\end{figure}

Fig.~\ref{fig.varl_varv} shows the dependence of $\mathrm{Im}(\omega)$ of the overstable mode on the viscosity parameters $l/r$ and $v_s/v_\phi$.
Fig.~\ref{fig.eigenfunction} shows $p_1(x)$ corresponding to the eigenfunction of the overstable mode for $\mathrm{Pr} = 0.2$, $k_r r = 40$, $l/r = 0.01$ and $v_s/v_\phi = 0.01$.

\section{Conclusions}
Using the anelastic approximation of linearized hydrodynamic equations, we studied the development of axially symmetric small perturbations in thin Keplerian discs.
Dispersion relation is derived as a solution of general Sturm--Liouville eigenvalue problem for pressure perturbations.
An overstable mode is discovered for different values of the disc thickness and microscopic viscosity and thermal conductivity of the gas.
The overstability appears when the Prandtl parameter exceeds a critical value $8/45$.
The unstable mode has viscous-convective nature and can serve as a seed for turbulence in astrophysical discs even in the absence of magnetic fields.

\section{Acknowledgement}
We thanks R.A.~Sunyaev for useful discussions. The work is supported by the Russian Foundation for Basic Research grant 14-02-91345.

\clearpage

\end{document}